\documentclass[12pt,aps,amsmath,latexsym,amsfonts]{JHEP3}
\usepackage{epsfig}
\usepackage{amsfonts,amssymb,amsmath}
\usepackage[hang]{subfigure}

\newcommand{\be}{\begin{equation}}
\newcommand{\ee}{\end{equation}}
\newcommand{\bea}{\begin{eqnarray}}
\newcommand{\eea}{\end{eqnarray}}
\newcommand{\beal}{\begin{aligned}}
\newcommand{\eeal}{\end{aligned}}

\newcommand{\Veff}{\ensuremath{V_{\text{eff}}}}
\def\c{{\rm c}}
\newcommand{\bse}{\begin{subequations}}
\newcommand{\ese}{\end{subequations}}

\def\tT{{\tilde T}}

\def\trho{{\tilde\rho}}
\def\tP{{\tilde P}}
\def\hphi{{\hat\phi}}
\def\hrho{{\hat \rho}}
\def\hP{{\hat P}}
\def\hV{{\hat V}}

\def\ms{M_*}
\def\rs{r_*}

\def\dphi{{\delta\phi}}

\def\R0{{R_0}}
\def\x1{x_\infty}

\def\as{\infty}

\title{Neutron Stars in Screened Modified Gravity: Chameleon vs Dilaton}

\author{Philippe Brax \\
  Institut de Physique Th\'eorique, CEA, IPhT, CNRS, URA 2306,
  F-91191Gif/Yvette Cedex, France \\ E-mail:
  \email{philippe.brax@cea.fr}}

\author{Anne-Christine Davis\\
  DAMTP, Centre for Mathematical Sciences, University of Cambridge,
  CB3 0WA, UK\\E-mail:
  \email{A.C.Davis@damtp.cam.ac.uk}}
\author{Rahul Jha\\
  DAMTP, Centre for Mathematical Sciences, University of Cambridge,
  CB3 0WA, UK\\E-mail:
  \email{r.jha@damtp.cam.ac.uk}}

\abstract{We consider the scalar field profile around relativistic compact objects such as neutron stars for a range of modified gravity models with screening mechanisms of the chameleon and Damour-Polyakov types. We focus primarily on inverse power law chameleons and the environmentally dependent dilaton as examples of both mechanisms. We discuss the modified Tolman-Oppenheimer-Volkoff equation and then implement a relaxation algorithm to solve for the scalar profiles numerically. We find that chameleons and dilatons behave in a similar manner and that there is a large degeneracy between the modified gravity parameters and the neutron star equation of state. This is exemplified by the  modifications to the mass-radius relationship for a variety of model parameters.}


\begin{document}
\maketitle
\flushbottom

\section{Introduction}

General Relativity is at the heart of our understanding of compact objects such as neutron stars \cite{Will2005}. This could be challenged by the observation of the
current accelerated expansion of the universe \cite{Clifton2011}.  Indeed, the cosmological acceleration has now been corroborated using multiple  probes \cite{Ade:2013zuv} since the original observations with supernovae \cite{Perlmutter1999,Riess1998} and whilst the addition of a cosmological constant term to Einstein's equations fits all the data extremely well there are appealing theoretical reasons to investigate alternative explanations \cite{Weinberg:1988cp}. One of these possibilities could be a modification of the laws of gravity \cite{Joyce:2014kja}.
As argued by Weinberg \cite{Weinberg:1965rz} any modification to the framework of general relativity necessarily introduces additional degrees of freedom. This leads to many infrared modified gravity theories which, in principle, could incorporate the late time acceleration of the universe. They all modify the gravitational interaction and this is bound by stringent experimental constraints in our local environment at least \cite{Bertotti2003, Williams2004}.  Generically, these alternative theories are valid when they include a screening mechanism that ``shields'' the effects of such  modifications to general relativity (see \cite{Joyce:2014kja} for a detailed review). Many proposals exist and they have all been extensively studied using laboratory
\cite{Mota2008,Brax:2008hh,Upadhye2012b,Upadhye2012,Upadhye2012a,Brax:2014zta,Brax:2011hb,Brax:2013cfa,Hamilton:2015zga,Burrage:2014oza,Brax:2016wjk,Lemmel:2015kwa}
solar system~\cite{Khoury2004}, cosmological~\cite{Jain2010}, and astrophysical~\cite{Jain2012} tests. These investigations all probe gravity in the weak field regime. With advances in experimental techniques, attention has increasingly
focussed on tests of gravity in the strong regime for compact objects such as neutron stars and black holes, \cite{Yunes2013,Psaltis2008}.
The effect of modified
gravity in the vicinity of astrophysical black holes has been
studied in \cite{Davis:2014tea,Davis:2016avf}

Many studies of the effects of modifying gravity in the strong gravity regime have originally  been limited to models subject to the chameleon screening mechanism.
For instance, there is a substantial number of works  on the nature of compact objects such as neutron stars in $f(R)$ gravity. For a long time it was claimed \cite{Frolov,KM1} that relativistic stars do not exist in these models due to the presence of an easily accessible singularity and other divergences springing from the functional form of $f(R)$ \cite{Frolov,Briscese,Abdalla}. Several exceptions are now known to exist, e.g.  \cite{BL1} gives the first  numerical construction of  static relativistic stars in $f(R)$ gravity. Other examples are known \cite{Tsujikawa, Upadhye-Hu} including numerical solutions corresponding to static stellar configurations with a strong gravitational field \cite{Babichev1, Babichev2}. More recently relativistic stars have been studied in Horndeski \cite{Cisterna:2015yla,Cisterna:2016vdx} and beyond Horndeski models \cite{Babichev:2016jom,Sakstein:2016oel,Sakstein:2016lyj}.

In this paper, we focus on screened modified gravity \cite{Khoury2010,Brax2012a,Brax2012} with two different screening methods: the chameleon \cite{Khoury2004,Brax2004a,Mota2007} and the Damour-Polyakov mechanisms \cite{Damour:1994zq,Brax2010,Hinterbichler2011}. We concentrate  on two typical examples. The first one is the inverse chameleon model which has already been analysed in the strong gravity regime and serves as a benchmark for our numerical results \cite{Babichev2}. The second one is the environmentally dependent dilaton \cite{Brax2010} which is motivated by string theory in the strong coupling regime and obeys the conjectured least coupling principle. It provides a simple and concrete example where the Damour-Polyakov mechanism is at play. We then apply the modified Tolman-Oppenheimer-Volkov equations due to the presence of the screened scalar field to these two models and study constant density and polytropic models of  neutron stars within the chameleon and dilaton theories. We also numerically compute the scalar profiles and mass-radius relationships for chameleons and dilatons  \cite{Capozziello:2015yza,Yazadjiev:2014cza,Yazadjiev:2015xsj}, and comment on parameter degeneracy for both models.

The outline of this paper is as follows. In section 2 we briefly review the screened modified gravity models that will concern us  and restrict ourselves to static, spherically symmetric spacetimes. In section 3 we discuss the Tolman-Oppenheimer-Volkoff equation and our numerical procedure. We  present our numerical results in section 4. We conclude with possible implications in section 5.

\section{Screened Modified Gravity}
\label{sec:modified}

\subsection{The models}

Our starting point is the Einstein-Hilbert action with the conformally  coupled scalar field in the Einstein frame,
\be
S=\int d^4 x \sqrt{-g}\left[\frac{M_{\rm Pl}^2}{2}R -\frac{1}{2}g^{\mu\nu}\partial_{\mu}\phi
\partial_{\nu}\phi -V(\phi)\right]+S_m\left[\Psi_i,A^2(\phi)g_{\mu\nu}\right].
\ee
Here $M_{\rm Pl}^2 = 1/8\pi G_N$ is the reduced Planck mass and $S_m$ is the action for matter fields (denoted generically as $\Psi_i$), which couple
minimally to the Jordan frame metric, $\tilde{g}_{\mu\nu} = A^2(\phi)g_{\mu\nu}$.
We assume $A(\phi)$ is close to 1 for values of $\phi$ in the allowed range below the Planck scale, i.e. we have
\be
\label{betadef}
A(\phi) = 1 + \int^\phi  d\tilde\phi \frac{\beta(\tilde \phi)}{M_{\rm Pl}}
\ee
where $A(\phi)$ and $\beta (\phi)$ will be specified later explicitly for both chameleons and dilatons.
As we are always in the low energy regime, we take $\phi \ll M_{\rm Pl}$ throughout.

The Einstein equation can be derived as
\be
G^{\mu\nu} = \frac{1}{M_{\text{pl}}^2}\left(T_m^{\mu\nu} +T_{\phi}^{\mu\nu}\right)
\ee
where the energy momentum tensor of the scalar field is given by
\be
T_{\phi}^{\mu\nu} = \nabla^{\mu}\phi\nabla^{\nu}\phi - g^{\mu\nu}\left(\frac{1}{2}g^
{\alpha\beta}\nabla_{\alpha}\phi\nabla_{\beta}\phi+V(\phi)\right)
\ee
and $T^{\mu\nu}_m$ is the Einstein frame energy-momentum of matter.
Defining the trace of the energy momentum of matter  $T_m \equiv g^{\mu\nu}T^m_{\mu\nu}$, the scalar field equation can
then be written as
\be
\square \phi
= \frac{\partial V}{\partial \phi} - \frac{\partial \ln A}{\partial \phi}T_m.
\ee
In particular we need to pay attention to  the conformal transformation between the Einstein frame metric $g_{\mu\nu}$ and the Jordan frame metric $\tilde{g}_{\mu\nu}$. The perfect fluid energy momentum tensor in the Einstein frame can be written as
\begin{equation}
T_{\mu\nu}=(\rho+P)u_{\mu}u_{\nu}+Pg_{\mu\nu}
\end{equation}
where \(u^{\mu}\) is the 4-velocity of the fluid elements, $\rho$ the energy density and $P$ the pressure. The Jordan frame energy-momentum tensor $\tT_{\mu\nu}$ and the Einstein frame energy momentum tensor, $T_{\mu\nu}$, are related by
\be
T^\mu_\nu = A^{4}\tT^\mu_\nu
\ee
so that, for a perfect fluid, we have
\be
\rho=A^4\trho, \quad P=A^4\tP
\ee
The trace of the energy momentum tensor is $T_m \equiv g^{\mu\nu}T^m_{\mu\nu} = -\rho + 3P$ in the Einstein frame.
The numerical values for the pressure and density are essentially the same in both frames given our constraint $\phi\ll M_{\rm Pl}$ and $A\sim 1$.

In the following we shall concentrate on two types of models subject to the chameleon and Damour-Polyakov mechanisms. Both types of scenarios require the existence of a minimum of the
effective potential $\phi(\trho,\tP)$
\be
V_{\rm eff}=V+\frac{1}{4} A^4 (\trho-3\tP)
\ee
in terms of the Jordan matter density and pressure, taken as conserved in the Jordan frame and scalar field independent.
For chameleons the effective mass of the scalar field at the minimum of the effective potential
\be
m_{\rm eff}^2(\trho,\tP) \equiv \frac{d^2 V_{\rm eff}}{d\phi^2}= \frac{d^2V}{d\phi^2}+(\frac{d\beta}{d\phi}+ 4\frac{\beta}{M_{\rm Pl}}) \frac{\beta}{M_{\rm Pl}} A^4(\phi) (\trho-3\tP)
\ee
must be  large enough in dense environments,  such as the interior of a star, in order to Yukawa suppress any effect of the scalar field on outside bodies. On the other hand, the Damour-Polyakov mechanism operates
when $\beta(\phi(\trho,\tP))$ becomes extremely small in dense environments and effectively suppresses all fifth force effects.

\subsection{Static and spherically symmetric configurations}

We are interested in the effect of screened modified gravity in the context of compact and relativistic objects. In such a case,
we can  restrict ourselves to  a static, spherically symmetric background geometry with
\be
ds^2=-e^{\nu(r)} dt^2+e^{\lambda(r)} dr^2+r^2 \left(d\theta^2+\sin^2\theta\, d\phi^2\right)
\ee
where the scalar field is such that $\phi = \phi(r)$ and we parameterise
\be
 e^{-\lambda(r)}= 1-\frac{2m(r)}{r}.
\ee
The $t$ and $r$ components of Einstein's equations then become,
\begin{align}
m' & =\frac{r^2}{2M_{\rm Pl}^2}\left[A^4 \trho+\frac12 e^{-\lambda}{\phi'}^2+V(\phi)\right] \label{eq5.36} \\
\nu'& =e^\lambda\left[\frac{2m}{r^2}+\frac{r}{M_{\rm Pl}^2}\left(\frac12 e^{-\lambda}{\phi'}^2-V(\phi)\right)+
\frac{rA^4 \tP}{M_{\rm Pl}^2}\right]\label{eq5.37}
\end{align}
where the fluid parameters are defined in the Jordan frame, i.e. we analyse the Einstein equations in the Einstein frame, where the gravitational constant is a true constant, as a function of the conserved fluid parameters as defined by the Jordan frame.
Therefore we use the conservation equation in the Jordan frame
\be\label{eq5.38}
\tilde\nabla_\mu \tT^{\ \mu\nu}_{m}=0
\ee
which gives us the relation
\be
\label{eq5.39}
\tP'=-\frac12\left(\trho+\tP\right)\left(\nu'+2\frac{A'}{A}\phi'\right).
\ee
The Klein-Gordon equation  also simplifies to
\be
\label{eq5.40}
\phi''+\left(\frac{2}{r}+\frac12(\nu'-\lambda')\right)\phi'=e^\lambda\left[\frac{dV}{d\phi}+A^3 A'(\trho-3\tP)\right].
\ee
As already stated, the source term on the right hand side can be seen  as an effective potential for the scalar field up to the metric component $e^\lambda$.
A  local extremum of the potential is characterised by
\be
\frac{d V_{\rm eff}}{d\phi}=\frac{dV}{d\phi}+\frac{\beta}{M_{\rm Pl}}A^4(\phi)(\trho-3\tP)=0.
\ee
The effective mass of the scalar field $\frac{d^2 V_{\rm eff}}{d\phi^2}$ at the minima  is always  positive if the potential $V(\phi)$ is convex and $(\trho-3\tP)>0$. As it turns out, the latter is an important condition for the stability of the solution.

\subsection{Stability}

A stability analysis of the scalar equation of motion~\ref{eq5.40} provides further insight. Following
~\cite{Harada:1997mr} we note that since we are in a static, spherically symmetric geometry, we can perturb the scalar into spherical harmonics ($Y_{lm}(\Omega)$) and ignore all other perturbations to leading order. This is enough for the qualitative argument presented here and this implies that
\begin{equation}
\delta\phi = \sum \delta\phi_{lm}(r)Y_{lm}(\Omega)e^{i\omega t}
\end{equation}
and the mode equation becomes
\begin{equation}
-\omega^2\dphi-e^{\nu-\lambda}\left[\dphi''+\left(\frac{\nu'-\lambda'}{2}+\frac{2}{r}\right)\dphi'\right]+
e^{\nu}\left[\frac{l(l+1)}{r^2}+m^2_{\rm eff}\right]\dphi=0.
\end{equation}
Focussing on the radial part, we get an approximate dispersion relation of the form
\begin{equation}
\omega^2 \simeq k^2 + m^2_{\rm eff}
\end{equation}
where $k$ is proportional to the inverse of the typical length scale in the problem, the stellar radius $r_\star$. For stability we require $\omega^2>0$. Using $\trho_\star\sim M_\star/ r_\star^3$ and $\Phi_\star\sim M_\star/(M_{\rm Pl}^2 r_\star)$, we end up with the condition
\begin{equation}
(1-3\tP/\trho)\Phi_\star \gtrsim-\frac{1}{24 \beta_0^2}
\end{equation}
where $\beta_0$ is the coupling in vacuum outside the object, $M_\star$ the stellar mass and $\Phi_\star$ the Newtonian potential at its surface. For neutron stars with $\Phi_\star\sim 0.1$ and $\beta_0 \sim 1$, we find that as soon as $(1-3\tP/\trho)\lesssim -1$
 we would generically expect instabilities to appear. Eventually, this would back react and  prevent the existence of the object itself.

The nature of this instability depends essentially on
one more  ingredient which characterises the compact object: the fluid equation of state, or more precisely the relation between pressure and matter density
\be
\label{eq5.41}
\tP=\tP(\trho)
\ee
which we specify in the Jordan frame and which closes the system of equations.
The equation of state deep inside a neutron star is not very well known. Several approximations exist and the uncertainties are reasonably large. For our purposes it will be sufficient to focus on a polytropic equation of state for the interior region of the neutron star. See \cite{EoSNS}--\cite{Lattimer:2006xb} for a detailed discussion. We thus use
\begin{equation}
\tilde \rho = \left(\frac{\tilde P}{K}\right)^{1/\Gamma} + \frac{\tilde P}{\Gamma-1}
\end{equation}
where $\Gamma$ is the polytropic index and $K$, of mass dimension $4(1-\Gamma)$, is a scale to be determined below. Realistic neutron stars can be parameterised by piecewise polytropic equations of state with $\Gamma \sim 1-3$. As it turns out, the lower the polytropic index, the stiffer the mass radius relationship. We use values close to $2$ for our numerical work.

\section{The scalar profile for chameleons and dilatons}

\subsection{Tolman-Oppenheimer-Volkoff equation}

We have to simultaneously solve equations~\ref{eq5.36}--\ref{eq5.41} and given their complexity  we will have to resort to numerics. We will come to the numerical implementation of our relaxation algorithm in the next section, but before we do that it is helpful to recall the solution in GR, i.e. when  neglecting the scalar back reaction.

The GR problem was solved by Tolman, Oppenheimer and Volkoff via what  is generally known as the Tolman-Oppenheimer-Volkoff (TOV) equation. It constrains the structure of a spherically symmetric body of isotropic material which is in static gravitational equilibrium, as modelled by GR. This equation is derived by solving Einstein's equations for a general time-invariant, spherically symmetric metric.
Let us consider an interior Schwarzschild metric which gives the following line element
\begin{equation}
ds^2 = -e^{\nu(r)}dt^2 + \left(1-\frac{2m(r)}{r}\right)^{-1}dr^2 + r^2d\Omega^2
\end{equation}
where $m(r)/G_N$ has the simple interpretation of being the mass inside the radius $r$ whilst $\nu(r)$ or more precisely $\nu(r)/2$ is known as the metric potential. In vacuum, outside the star, $m=G_N M_\star$, $T_m^{\mu\nu} = 0$ and $\nu(r) = \ln (1-2G_N M_\star/r)$ where $M_\star$ is the total stellar mass.
Now assuming a barotropic equation of state (i.e. the pressure is a function of density alone), the system is governed by the set of equations which can be obtained by discarding the scalar field from the equations derived previously
\begin{align}
\tilde P' &= -(\tilde \rho + P) \frac{m+r^3\tilde P/2M_{\rm Pl}^2}{r^2(1-2m/r)} \label{eq5.45}\\
m' & =\frac{r^2}{2M_{\rm Pl}^2}\tilde \rho \\
\nu'& =\left(1-\frac{2m}{r}\right)^{-1}\left[\frac{2m}{r^2}+\frac{r\tilde P}{M_{\rm Pl}^2}\right]
\end{align}
where conventionally eq~\ref{eq5.45} is known as the TOV equation.

Using for the density a constant $\tilde \rho = \tilde \rho_0$, this leads to
\begin{equation}
m(r)=\frac{\tilde \rho_0 r^3}{6 M_{\rm Pl}^2}
\end{equation}
and one gets for the pressure
\begin{equation}
\tilde P(r)=\tilde \rho_0 \frac{\left(1-{\frac{2G_N\ms}{r_\star}}\right)^{1/2}-\left(1-{\frac{2G_N\ms r^2}{r_\star^3}}\right)^{1/2}}{\left(1-{\frac{2G_N\ms r^2}{r_\star^3}}\right)^{1/2}-3 \left(1-\frac{2G_N\ms}{r_\star}\right)^{1/2} }\, .
\end{equation}
The trace of the energy momentum tensor can be computed as,
\be
\tilde \rho-3 \tilde P=  \tilde \rho_0 \frac{4\left(1-{\frac{2G_N\ms r^2}{r_\star^3}}\right)^{1/2}-6\left(1-{\frac{2G_N\ms}{r_\star}}\right)^{1/2}}{\left(1-{\frac{2G_N\ms r^2}{r_\star^3}}\right)^{1/2}-3 \left(1-{\frac{2G_N\ms}{r_\star}}\right)^{1/2} }.
\ee
This quantity is of great interest as it directly couples to the scalar field which is here considered as a test field.  As  $\tilde \rho-3 \tilde P$ is an increasing function of $r$ , we find  its minimal value at the centre of the star
\be
\tilde \rho_c-3 \tilde P_c=\tilde \rho_0 \frac{4-6\left(1-2\Phi_\star\right)^{1/2}}
 {1-3 \left(1-2\Phi_\star\right)^{1/2}}
\ee
where $\tilde\rho_c$ and $\tilde P_c$ are respectively the density and pressure at the stellar centre and $\Phi_\star=G_N\ms / \rs$ is the gravitational potential at the stellar surface. Thus if $\Phi_\star$ is greater that $5/18$,  we can see that $\tilde \rho-3\tilde P$ is  negative in the stellar centre  and this potentially causes an instability in the scalar equation of motion \cite{Babichev1,Babichev2}. When this is not the case, one needs to resort to numerical results in order to understand the full dynamics. In the following, we shall concentrate on inverse power law chameleons and environmentally dependent dilatons for which the dynamics will be analysed.

\subsection{Chameleons}

\begin{figure}[!htb]
\centering
\includegraphics[scale=0.4]{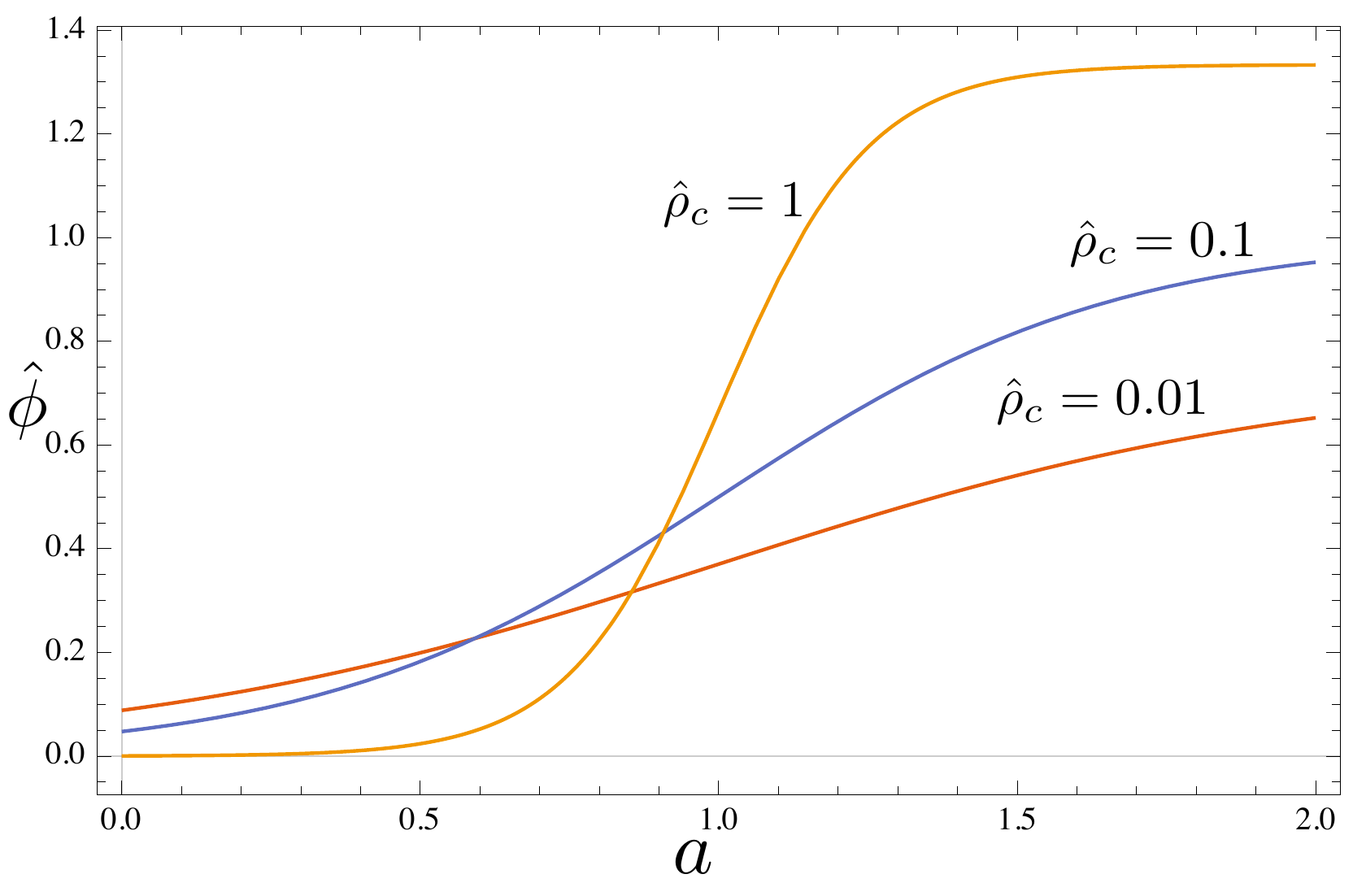}
\includegraphics[scale=0.4]{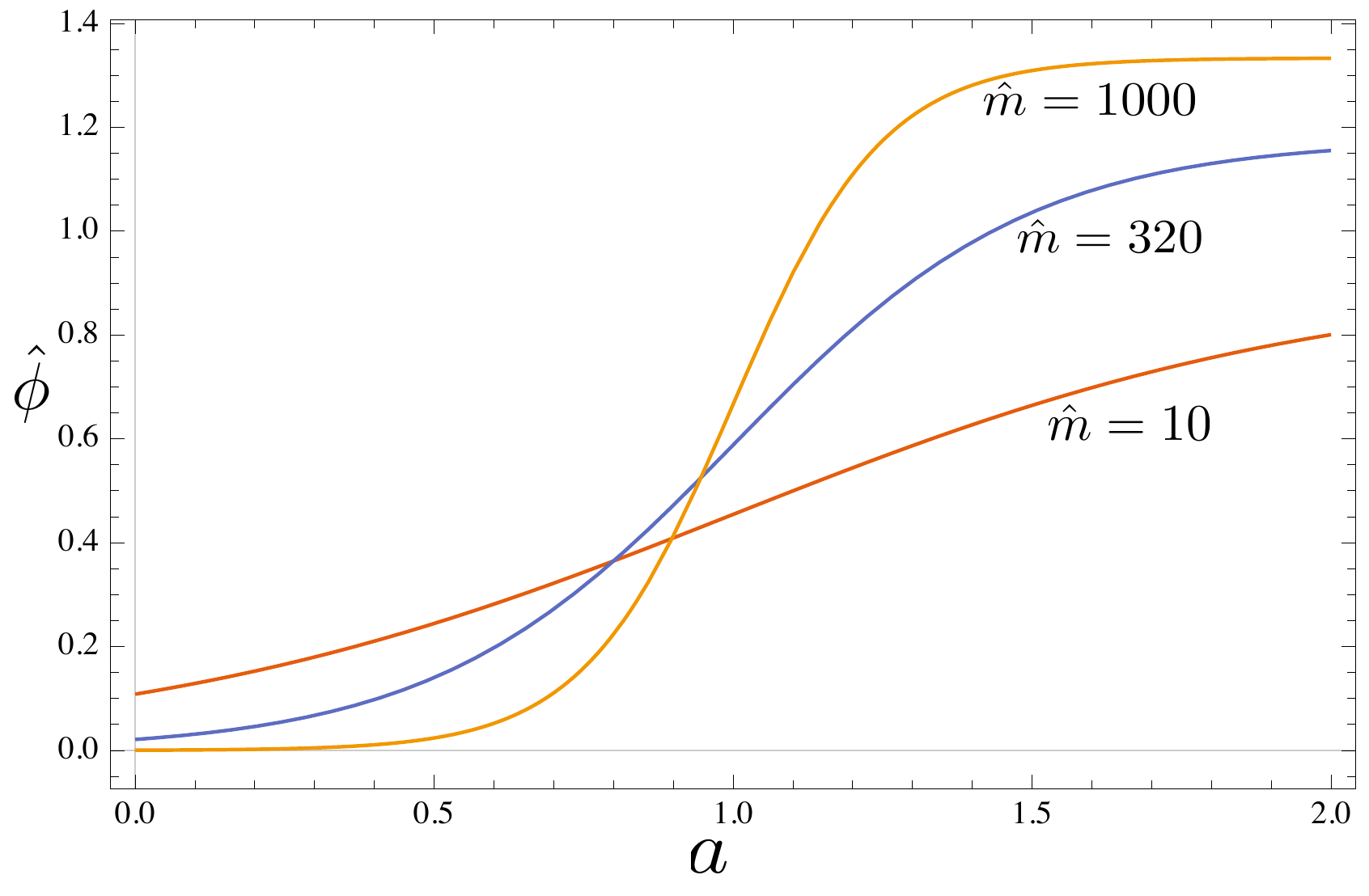}
\caption{In the left panel,  numerical solutions for the chameleon field for a
range of values of the stellar central density for a constant density star with  $M_0=0.01 M_{\rm Pl} $, $\hat m = 100$. In the right panel, the chameleon field for a range of values of the chameleon mass for a constant density star with  $\hrho_c = 0.1$,  $M_0=0.01 M_{\rm Pl} $.}
\label{chamstar1}
\end{figure}

The inverse power law chameleons combine an inverse power law potential for a scalar field with a constant coupling to matter \cite{Khoury2004}. It is well known that the chameleon mechanism works for these models and that the local tests of gravity impose a constraint on the scale $M$ appearing in the scalar potential, which can be taken of the order of the dark energy scale \cite{Khoury2004}. The coupling function is simply given by
\be
A(\phi) = e^{\beta\phi/M_{\rm Pl}}\,.
\ee
where $\beta$ is constant. The interaction potential is
\be\label{chV1}
V(\phi) = M^{4+n} \phi^{-n}=  V_0 \phi^{-n},
\ee
where $n\geq 1$ is an integer of order one, and we define
$V_0\equiv M^{4+n}$ to simplify notation. Keeping only the leading
order term from the coupling function, we
see that the effective potential is
\be
\Veff(\phi,\rho)\approx \frac{V_0}{\phi^{n}} + \frac{\beta\phi}{M_{\rm Pl}}(\trho - 3\tP).
\ee
Numerically, we first focus on  the simple case of a constant density star with an equation of state $\tilde \rho = \tilde \rho_0$. For numerical ease we use a smoothed  out ($1-\tanh$) transition to the asymptotic matter distribution surrounding the star. This can either be vacuum or a constant density Schwarzschild de Sitter geometry. Realistic stars should require matching to the time dependent cosmological background geometry, however our solution should capture all the salient features of this more general case.

\subsection{Dilatons}

We follow a similar procedure for the dilaton fields. The environmentally dependent dilaton \cite{Brax2010} is such that the coupling function has a minimum at $\phi=\phi_d$ as prescribed by the least coupling principle. The potential is taken to be exponentially decreasing as conjectured for the string dilaton in the strong coupling regime. The screening behaviour of dilatons is a consequence of the fact that their coupling function has a minimum at $\phi_d$,
\be
A(\phi) = 1+\frac{a_2}{2M_{\rm Pl}^2}(\phi - \phi_d)^2 +\dots
\ee
or equivalently we have for the coupling
\be
\beta(\phi) \equiv M_{\rm Pl} \frac{\partial \ln A(\phi)}{\partial \phi}
\thickapprox \frac{a_2}{M_{\rm Pl}} (\phi-\phi_d)\,.
\label{dil_betadef}
\ee
The dilaton self-interaction potential is
\be
V(\phi) =A^4(\phi)M^4 e^{-\frac{\phi}{M_{\rm Pl}}} \equiv A^4(\phi)
V_0 e^{-(\phi-\phi_d)/M_{\rm Pl}}
\ee
where again we define $V_0\equiv M^4 e^{-\phi_d/M_{\rm Pl}}$ to simplify notation.
The effective potential is thus
\begin{equation}
\Veff(\phi,\tilde \rho) = A^4(\phi)V_0 e^{-(\phi-\phi_d)/M_{\rm Pl}}+(\trho - 3\tP)(A(\phi) - 1)
\end{equation}
which can be  expanded  in $\phi/M_{\rm Pl}$. Solar system tests of the dilaton models imply that typically one must have $a_2\gtrsim 10^6$ \cite{Brax2010} guaranteeing that the mass
of the dilaton, on large cosmological scales, is larger than $10^3 H_0$ \cite{Brax2012}, where $H_0$ is the Hubble rate now.

\subsection{Numerical procedure for neutron stars}

The solution of the equations of motion given a specified equation of state requires a numerical procedure.
We solve the above set of equations~\ref{eq5.36}--\ref{eq5.40} together with the  equation of state numerically. We follow the general framework of~ \cite{Babichev:2009fi} where the authors study relativistic stars  in $f(R)$ gravity. In our case  we expect the scalar field to exhibit a smoothed out kink-like behaviour extrapolating between the minimum inside the neutron star to the minimum of the ambient matter density outside the stellar surface.  The  boundary conditions are set at $r=0$ and not at the surface of the star $r=r_\star$. For numerical ease we begin by rescaling all the variables
\be
\label{rescaling}
a=\frac{r}{r_0}, \ \
\hphi = \frac{\phi}{M_0}, \ \
\hrho =\frac{r_0^2}{M_{\rm Pl}^2}\trho, \ \
\hP =\frac{r_0^2}{M_{\rm Pl}^2}\tP
\ee
where $r_0$, $M_0$ are  parameters specific to the star/modified gravity theory being considered. We will make them explicit later.

Substituting the above variables into the Einstein and scalar equations of motion we obtain,
\begin{align}\label{eq5.46}
\frac{\left(a\left(1-e^{-\lambda}\right)\right)'}{a^2}&= \hrho e^{4\beta M_0\hphi/M_{\rm Pl}}  + \left(\frac{M_0}{M_{\rm Pl}} \right)^2\left(\frac12 e^{-\lambda}{\hphi}'^2+V\frac{r_0^2}{M_0^2}\right) \\ \nonumber
\frac{\nu' e^{-\lambda(r)}}{a}&= \frac{\left(1-e^{-\lambda}\right)}{a^2}+ \hP e^{4\beta M_0\hphi/M_{\rm Pl}} + \left(\frac{M_0}{M_{\rm Pl}} \right)^2\left(\frac12 e^{-\lambda}{\hphi}'^2-V\frac{r_0^2}{M_0^2}\right)\\ \nonumber
\hP' &=  -\frac12\left(\hrho+\hP\right)\left(\nu'+2 \beta \frac{M_0}{M_{\rm Pl}} \hphi'\right),\\ \nonumber
0 &=  \hphi''+\left(\frac{2}{a}+\frac12(\nu'-\lambda')\right)\hphi'
- e^\lambda\left[\frac{d\hV}{d\hphi}+\frac{M_{\rm Pl}}{M_0}\beta e^{4\beta M_0\hphi/M_{\rm Pl}}(\hrho-3\hP)\right]
\end{align}
The parameter $M_0$ essentially sets a mass scale for the scalar field and will depend on the modified gravity theory considered.
To close this system of equations, we also need the equation of state for the fluid making up the star and written out in terms of the rescaled variables this reads,
\be
\label{eos-r}
\hP=\hP(\hrho).
\ee
We use a polytropic equation of state and in our rescaled variables we get
\begin{equation} \label{eq5.67}
\hrho = \left(\frac{\hP}{\hat K}\right)^{1/\Gamma} + \frac{\hP}{\Gamma-1}
\end{equation}
$\hat K$ here is the overall normalisation of the polytropic equation of state and for us it is just a numerical constant. We pick the value of $0.39$ which is consistent with what is used in the literature~\cite{Ozel:2009da}.
We now need to pick a value for the scale $r_0$ used to normalise all length scales in our problem. A convenient choice is of the order of
\begin{equation}\label{r0}
r_0=M_{\rm Pl} \trho_\c^{-1/2}
\end{equation}
so that for $\Gamma = 2$ and $\hat K=0.39$ we  get $\hrho_\c=1$  and  $\hP_c \simeq 0.23$. With these choices $r_* \simeq r_0 \times \mathcal O(1)$ for highly relativistic stars, where $r_*$ is the stellar radius.

The choice of the various numerical parameters is  motivated by fixing the normalisation scale to a specific, stable,  stellar solution of the Tolman-Oppenheimer-Volkoff equations in GR. It can be shown \cite{Babichev:2009fi} that such a solution  corresponds to a star with central energy density of $\tilde{\rho}_c = 6.47 \times 10^{14} \;\rm{g/cm}^3$, a global mass of $M_\star = 3.10 M_{\odot}$ and a gravitational potential of $|\Phi_\star|\simeq 0.21$. The stellar configuration in chameleon and dilaton gravity theories is almost identical to this solution.
We use a relaxation type algorithm and vary the relaxation parameter to ensure fast and smooth convergence. We vary the grid size by hand to zoom in on regions where the scalar field varies rapidly. A non-homogeneous grid would provide better resolution but is beyond the scope of this work. We do however note that for the theories and parameter ranges considered, there is no significant loss of resolution.

We now turn to the  boundary conditions. We have three first-order differential equations and one second-order differential equation and thus we require five boundary conditions.
\begin{enumerate}
\item Demanding regularity at the centre of the star ($a=0$) implies
\begin{align}
\hphi'(0)&=0\\ \nonumber
(a(1-e^{-\lambda}))|_{a\rightarrow 0} &= 0
\end{align}
where the second condition ensures that the metric functions  are well behaved at the stellar centre. This corresponds to a vanishing mass function $m(r)$ at the origin.

\item The condition on $\nu$ corresponds to defining the time coordinate as the proper time of a static observer at a point $a_1$ far from the stellar surface.
 \begin{equation}
 \nu(a_1)=0
 \end{equation}

\item As $a_1\rightarrow \infty$ we expect
\be
\hphi(a_1)|_{a_1 \rightarrow \infty}=\hphi_\as
\ee
This ``asymptotic'' value $\hphi_\as$ corresponds to the minimum of the effective potential $V_{\rm eff}$ far from the star where this behaviour can be solved for analytically. In the limit of $a\rightarrow\infty$ the scalar equation becomes
\begin{equation}
\hphi''+\frac{2}{a}\hphi' = \frac{d\hat V}{d\hphi}
\end{equation}
and for a massive scalar field the solution looks like
\begin{equation}
\hphi = \hphi_{\infty}+B\frac{e^{-\hat m a}}{a}
\end{equation}
where $\hat m$ is the scalar mass and $B$ is a constant. Numerically it is easier to use this fall off rather than imposing a hard boundary condition.
\item The boundary conditions for the pressure are either
\begin{align}
\hP(0)&=\hP_c\\ \nonumber
\hP(a_1)&=\hP_\infty  \nonumber.
\end{align}
\end{enumerate}
For a constant density star we choose $\hP(a_1) = -\hrho_{\infty}$ so that the solution is asymptotically de Sitter, while for the polytropic equation of state it is easier to fix the pressure at the stellar centre and this once again gives an asymptotically de Sitter solution.

\section{Numerical Results}
In this section we present the results of our numerical integration. We carry out the integration for chameleon and dilaton modified gravity models.

\subsection{Chameleons}

For fixed values of the coupling $\beta \sim 1$ and the parameter $n = 1$, the numerical solution changes with the central density $\hrho_c$, the mass scale $M_0$ and the asymptotic density $\tilde \rho_\infty$.
More concretely, we take the density to vary according to
\begin{equation} \label{eq6.52}
\hrho(a) = \frac{\hrho_c}{2}\left[1-\tanh \left( \frac{a-1}{\epsilon} \right)\right]+\hrho_{\infty}.
\end{equation}
Choosing $\epsilon$ to be sufficiently small we recover the step function profile where $\hrho \sim \hrho_c$ for $a \lesssim 1$ and $\hrho \sim \hrho_{\infty}$ for $a \gtrsim 1 $ and thus the stellar radius is around $a \sim 1$. The $\hrho_{\infty}$ value should be  the typical galactic density  $\tilde \rho_{\infty} \sim 10^{-24} \rm{g/cm^3}$. This implies that $\hat \rho_\infty \sim 10^{-38}$. This is a  more appropriate asymptotic density than typical cosmological densities as we expect neutron stars to be in galaxies and as is apparent from Eq. \eqref{eq6.52} for small enough values of $\epsilon$, $\hrho \sim \hrho_{\infty}$ for $a \gtrsim 1 $. We do however note, that given the large density contrast between the neutron star interior and the asymptotic density used, we do not find our numerics to be very sensitive to this parameter. For numerical convenience we will use a larger value of $\hat \rho_{\infty} \sim 10^{-4}$.   Realistic stars will require matching to the time dependent cosmological background geometry, however our solution should capture all the salient features of this more general case.

We always set $M_0$ such that the asymptotic value of the scalar field is normalised to a value $\mathcal{O}(1)$.
This is done purely for numerical ease and in particular we use the convenient value of
\be
M_0 = \frac{1}{2.6}\sqrt{\frac{V_0M_{\rm Pl}}{\beta \trho_{\infty}}}.
\ee
Thus the asymptotic value of the scalar field is given by
\be
\hphi_{\infty} \simeq 2.6 \left(\frac{\trho_{\infty}}{\trho-3\tP}\right)^{1/2}
\ee
so that  we have $\hphi_{\infty} \sim 1.3$. We define the  rescaled potential as
$
\hat{V} = V\frac{r^2_0}{M_0^2}
$
expressed in terms of the rescaled variables as
\be\label{veff11}
\hat{V}_{\rm eff} = \frac{\hat{V}_0}{\hphi} +\frac{M_0^2}{4M_{\rm Pl}^2} e^{4\beta M_0\hphi/M_{\rm Pl}}(\hrho - 3\hP)
\ee
and the effective mass as
\be \label{meffective}
\hat{m}^2_{\rm eff} = \frac{2\hat{V}_0}{\hphi^3_{\rm min}} + 4\beta^2 e^{4\beta M_0\hphi_{\rm min}/M_{\rm Pl}}(\hrho -3\hP ).
\ee
For our numerics, we use the central density of the star to define the effective mass. Note that all of this is defined up to an overall numerical parameter $\hat{V}_0$. Realistic values for the chameleon with $n=1$ are such that $V_0 \sim 10^{-15}$ eV. This is far too small a number for numerics and we instead use $\hat{V}_0 \sim \mathcal{O}(0.01)$.

The effective rescaled mass is large in the stellar interior and small outside and thus we  expect the scalar to track the local minima of the effective potential in the stellar interior and smoothly evolve towards $\hphi_{\infty}$ outside the star. We can further see that the sign of $(\hrho -3\hP)$ is crucial in determining if this effective mass is well defined and consequently if the solution is stable.

We plot the scalar field profile for a range of central densities $\hrho_c$ in the left panel of figure ~\ref{chamstar1}. For low central densities the field smoothly transitions from its minima inside the star to the asymptotic minimum far from the stellar surface. Increasing this scaled central density while keeping the radius fixed has the effect of increasing the physical density of the star and this in turn means that the field is very stiffly held on to its minimum value inside the star. There is now a sharp transition close to the surface of the star, much like the chameleon thin shell effect. The field then smoothly approaches its asymptotic global minimum outside the star. It is interesting to note that this behaviour is almost exactly what we get if we keep the central density of the star fixed and  tune the effective mass of the chameleon scalar inside the star. This is shown in the right panel of figure~\ref{chamstar1}. There are two complementary explanations for this behaviour. In the first case it is easier to think in terms of stellar parameters. Our relaxation algorithm tries to minimise the net energy (potential plus gradient) with our boundary conditions of $\hphi'=0$ at $a=0$ and $\hphi$ fixed asymptotically. As the energy density inside the star increases it is increasingly unfavourable for the scalar to deviate significantly from its minimum value. Close to the stellar surface the smoothed out kink in $\rho$ forces the scalar to quickly evolve to its asymptotic value. Changing the mass of the scalar has a similar effect, a large mass, i.e. a small Compton wavelength, scalar  switches between the two minima sharply whereas a small mass, large Compton wavelength, scalar has a sluggish transition.
\begin{figure}[!htb]
\centering
\includegraphics[scale=0.4]{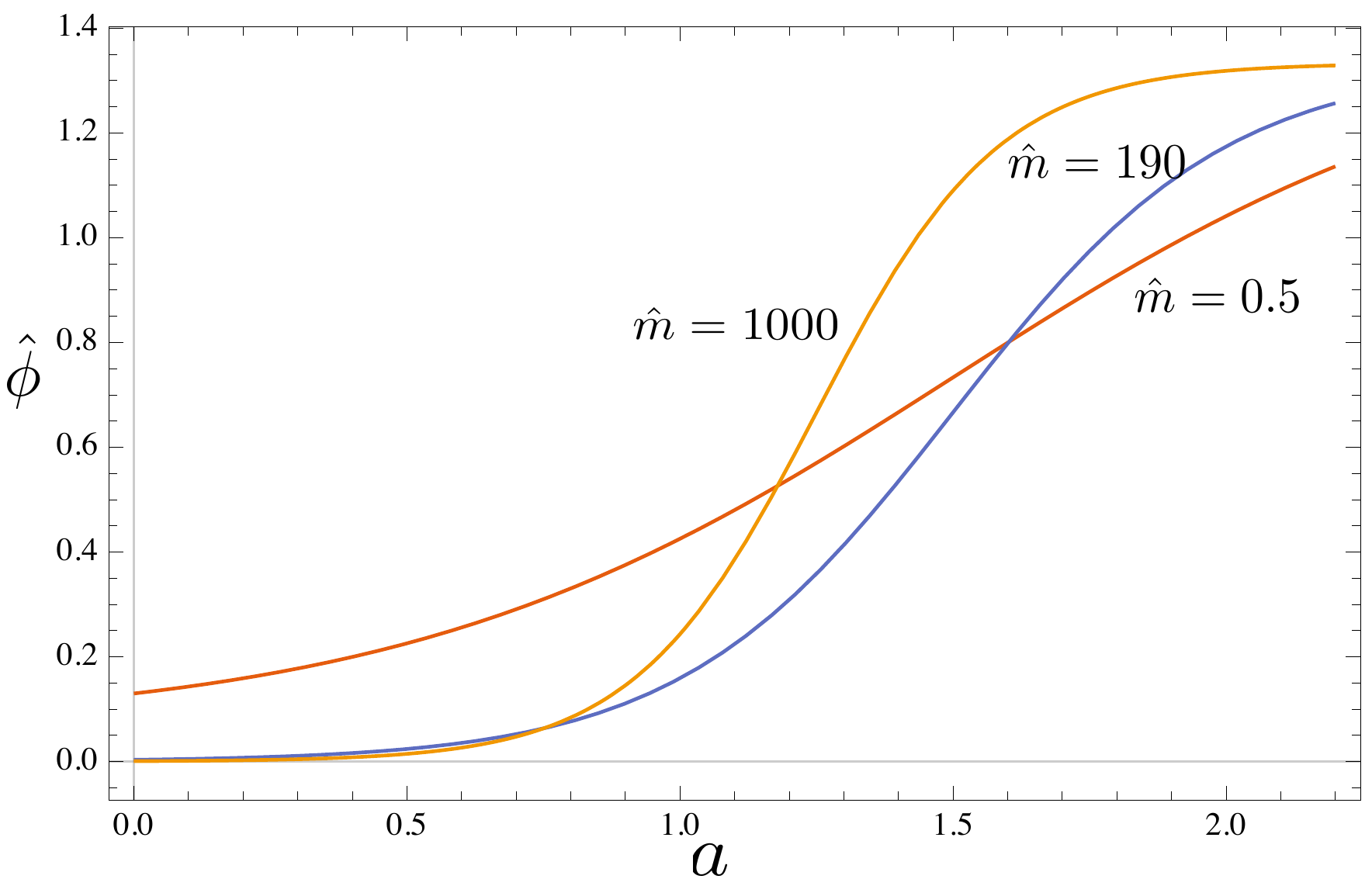}
\includegraphics[scale=0.4]{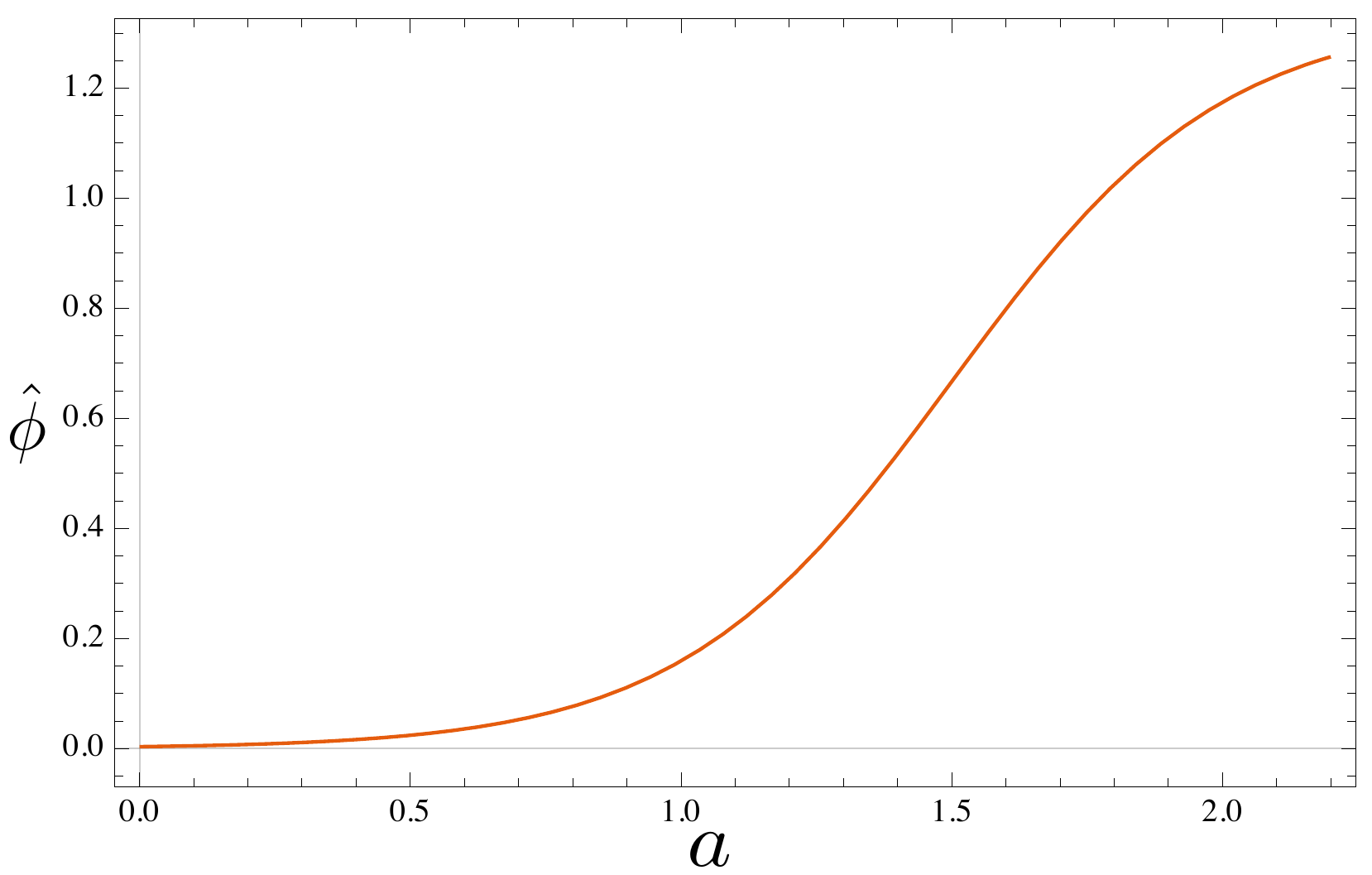}
\caption{In the left panel, the chameleon field for a range of values of the chameleon mass and a polytropic neutron star. The parameters used here are $\Gamma = 2$, $\hP_c = 0.23$, $\hrho_{\infty} = 10^{-4}$, $M_0=0.01 M_{\rm Pl} $. In the right panel
the chameleon field with $\Gamma = 2.12$ which  mimics the behaviour of the $\hat{m} = 190$ in the right panel even though the parameters are  $M_0=0.01 M_{\rm Pl} $, $\hat m = 100$ and $\hrho_{\infty} = 10^{-4}$}
\label{chamstar3}
\end{figure}

For a polytropic equation of state for the neutron star we get a qualitatively similar behaviour. A slight modification to eq.~\ref{eq5.67} means that we can smoothly transition to the ambient background/cosmological density and this also helps us avoid certain numerical instabilities close to the stellar surface. This point is discussed in greater detail in \cite{Babichev:2009fi}. We thus use
\begin{equation}\label{eq5.77}
\hrho = \left(\frac{\hP+\hrho_{\infty}}{0.39}\right)^{1/\Gamma} + \frac{\hP}{\Gamma-1}
\end{equation}
where $\hrho_{\infty}$ is the ambient density outside the star and is equal to zero for vacuum. The left panel in figure ~\ref{chamstar3} shows the evolution of the scalar field for a range of  scalar masses respectively. The right panel shows that similar profiles for different values of the central masses can be obtained by changing $\Gamma$ respectively. Hence there is a strong  degeneracy between the neutron star equation of state and the modified gravity model parameters.

\begin{figure}[!htb]
\centering
\includegraphics[scale=0.4]{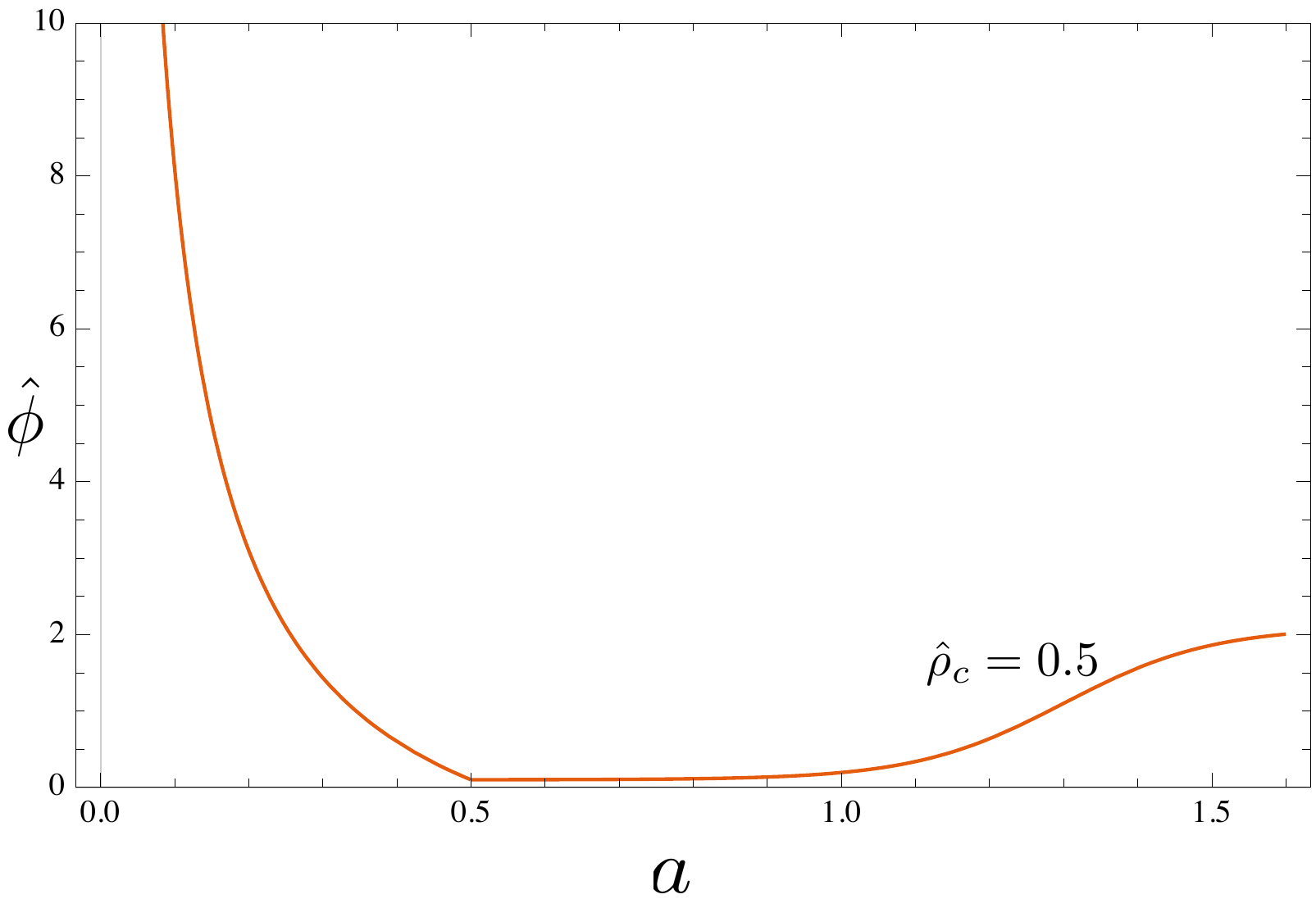}
\includegraphics[scale=0.4]{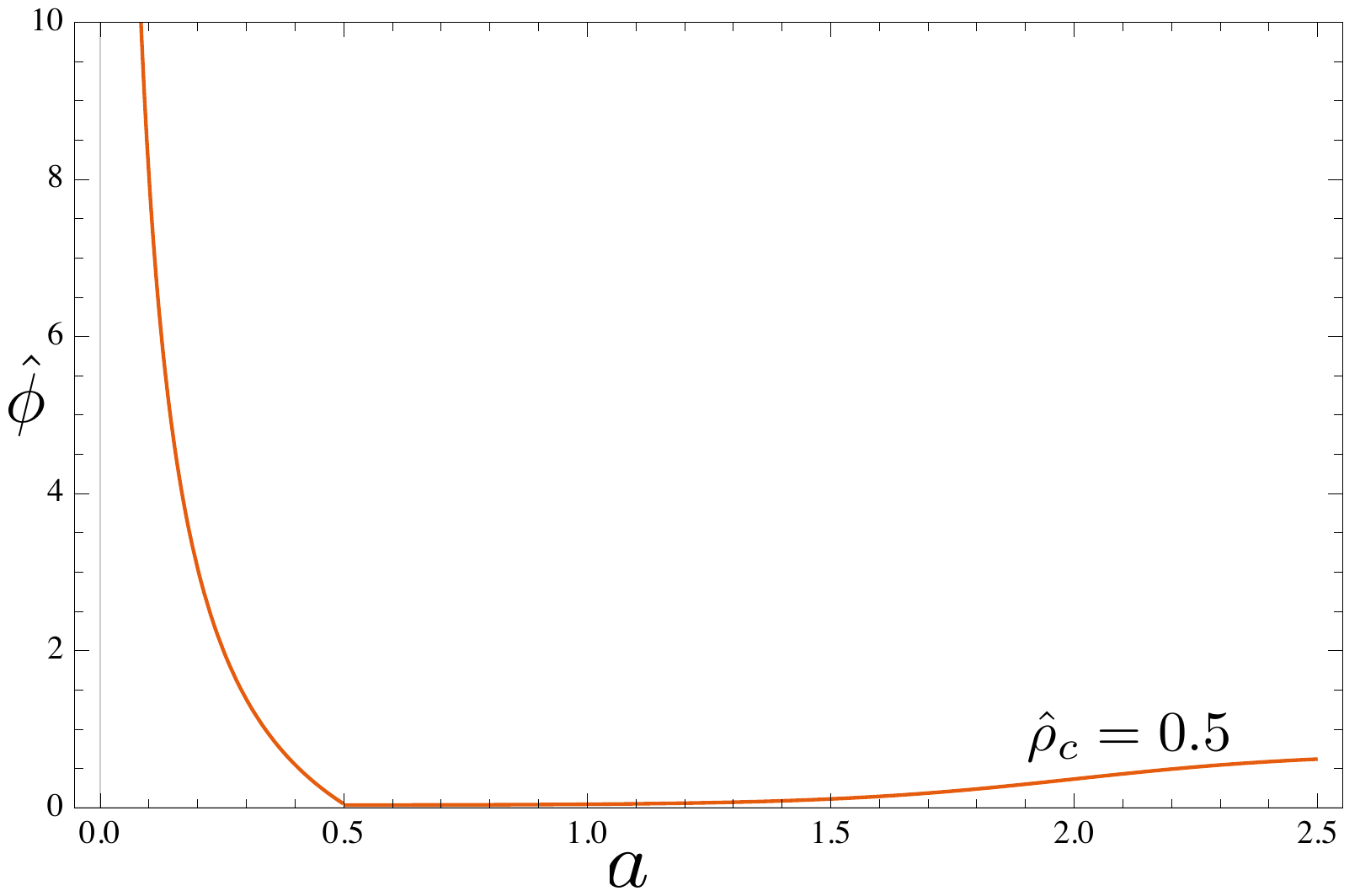}
\caption{In the left panel, we show an example of  unstable chameleon solution whilst in the right panel we have an unstable dilaton.}
\label{chamstar5}
\end{figure}

Finally for the sake of completeness we plot the scalar field profile for the case of $\hrho-3\hP<0$ in the centre of the star in the left panel of fig.~\ref{chamstar5}. As mentioned previously, in this case the scalar field does not have a well defined minimum inside the star and diverges. This divergence obviously means that the system is no longer stable as the energy momentum tensor associated with the scalar also diverges implying that the background geometry is no longer stable.

\subsection{Dilaton}

In the case of dilatons, we get the following expansion in $\phi/M_{\rm Pl}$ for the effective potential
\begin{equation}
V_{\rm eff}(\phi, \rho) \simeq V_0-V_0\left(\frac{\phi - \phi_d}{M_{\rm Pl}} \right)+\frac{1}{2}\left(\frac{\phi - \phi_d}{M_{\rm Pl}} \right)^2 a_2(4V_0+(\trho-3\tP))+...
\end{equation}
The minimum of the effective potential is obtained  at a value $\phi_{\rm min}$ defined by
\begin{equation}
(\phi_{\rm min} - \phi_d) = \frac{M_{\rm Pl}V_0}{a_2(4V_0+(\trho-3\tP))}.
\end{equation}
As for chameleons  we rescale the  field so that at infinity when $\tP_{\rm \infty}/\trho_{\rm \infty}= -1$ we get a rescaled field of order $\mathcal{O}(1)$. Moreover, it is convenient to shift and redefine $(\phi - \phi_d)\to \phi $. Using the scale
\begin{equation}
M_0 = \frac{V_0M_{\rm Pl}}{2a_2(V_0+\trho_{\infty})}
\end{equation}
we obtain that
\begin{equation}
\hphi_{\rm min} = \frac{2(V_0+\trho_{\infty})}{4V_0+(\trho-3\tP)}
\end{equation}
and at infinity, $\hphi_{\infty} = 0.5$.
We can now express  the rescaled effective potential $\hat{V}_{\rm eff}$  in terms of the rescaled variables
\begin{align}
\hat{V}_{\rm eff}(\hphi,\hrho) &= \frac{r_0^2}{M_0^2}\left[A^4(\phi)V_0 e^{-\phi/M_{\rm Pl}}+\frac{1}{4}A^4(\phi)(\trho - 3\tP)\right]\\ \nonumber
&= e^{4\beta(\hphi) M_0\hphi/M_{\rm Pl}}\left[\hat{V}_0 e^{-M_0\hphi/M_{\rm Pl}}+\frac{M_0^2}{4M_{\rm Pl}^2}(\hrho - 3\hP)\right]
\end{align}
and derive the corresponding mass.
 It should be noted that these expressions are defined up to an overall numerical constant $\hat{V}_0$ which we take to be $\mathcal{O}(0.01)$ for numerical ease.

\begin{figure}[!htb]
\centering
\includegraphics[scale=0.4]{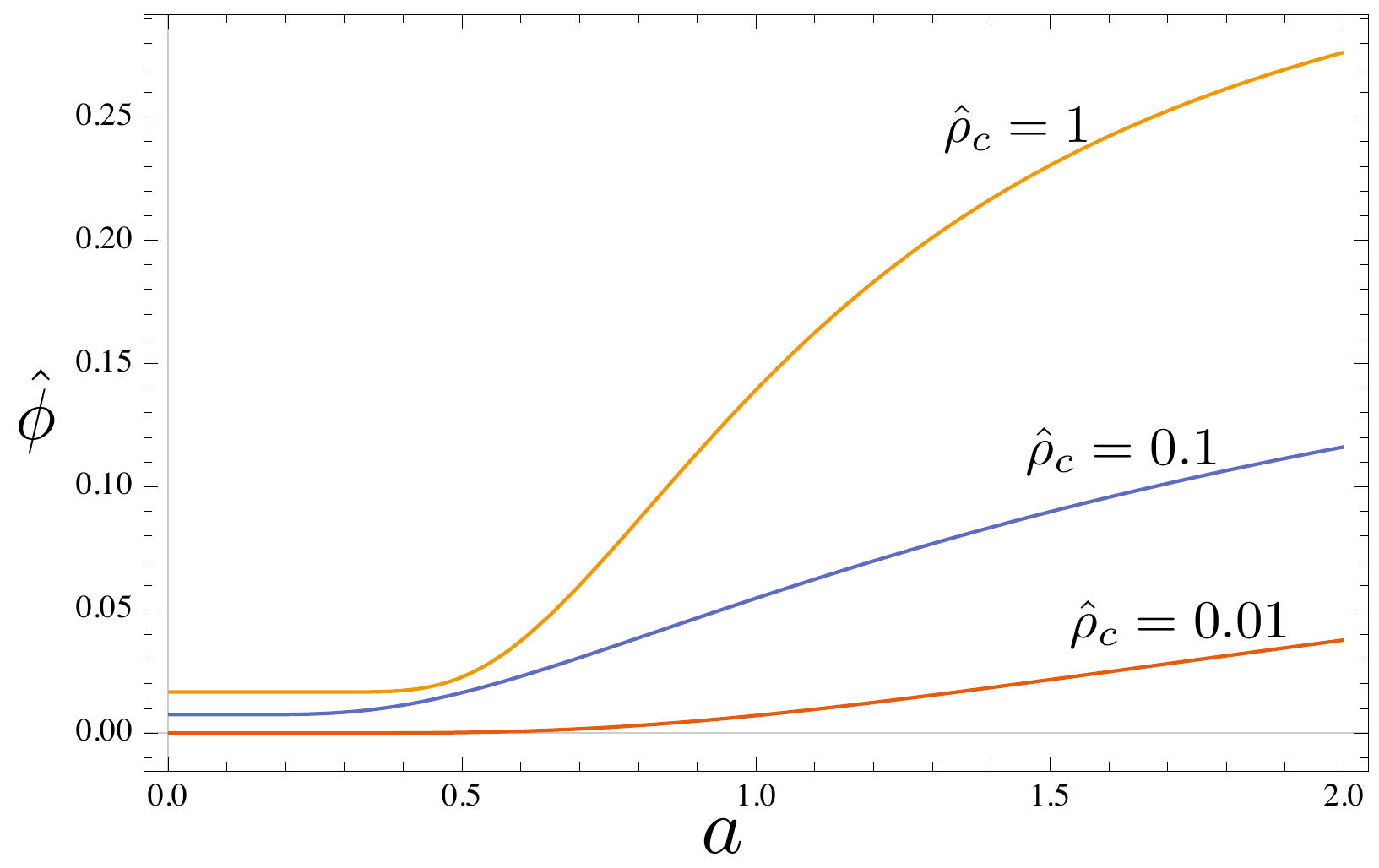}
\includegraphics[scale=0.4]{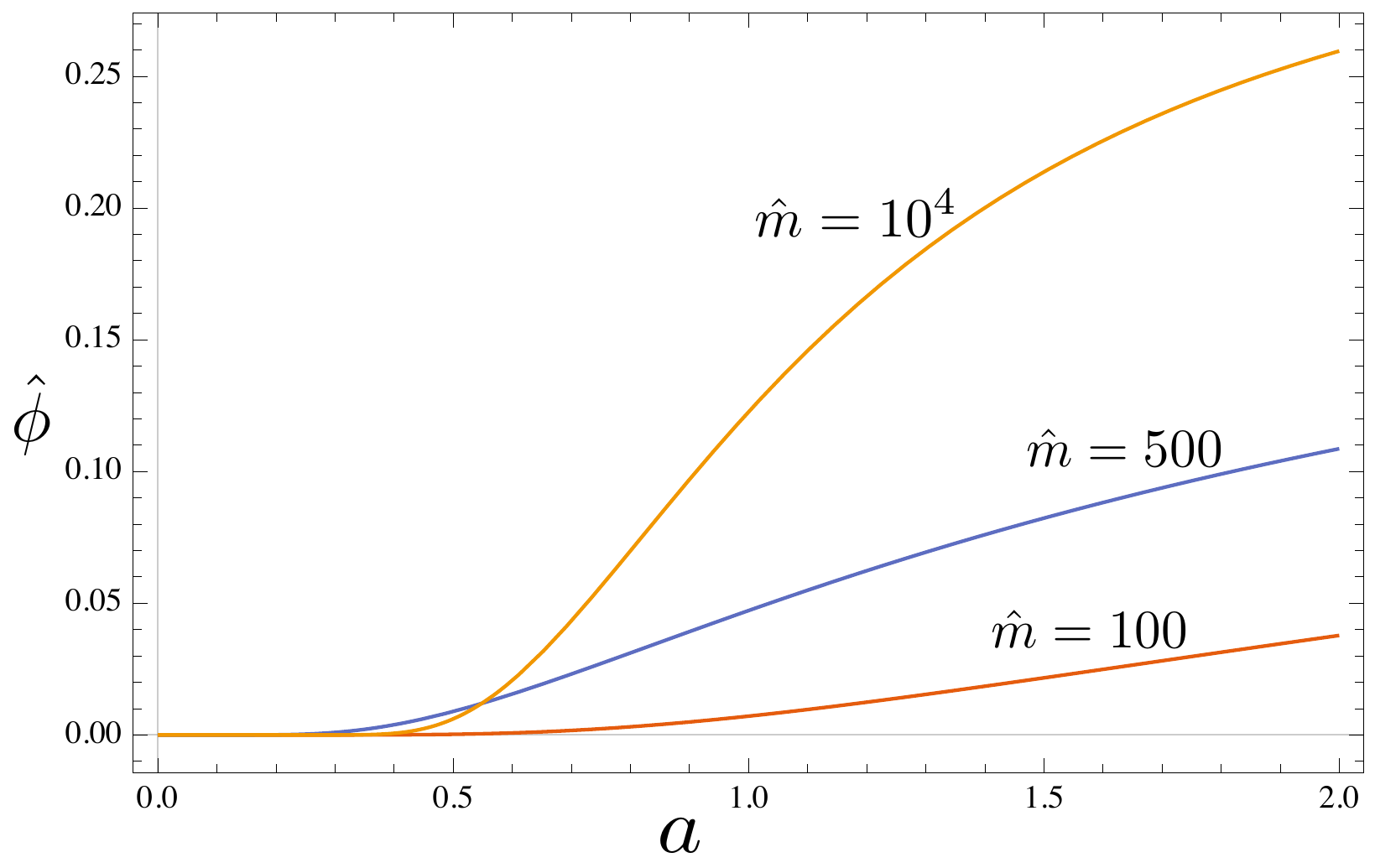}
\caption{In the left panel,  numerical solutions for the dilaton field for a
range of values of the stellar central density and a constant density star with $M_0=0.01 M_{\rm Pl} $ and $\hat m = 500$.
In the right panel, the dilaton field for a
range of values of the dilaton mass and a constant density star with  $\hrho_c = 0.1$ and $M_0=0.01 M_{\rm Pl} $.
}
\label{dilstar1}
\end{figure}

\begin{figure}[!htb]
\centering
\includegraphics[scale=0.4]{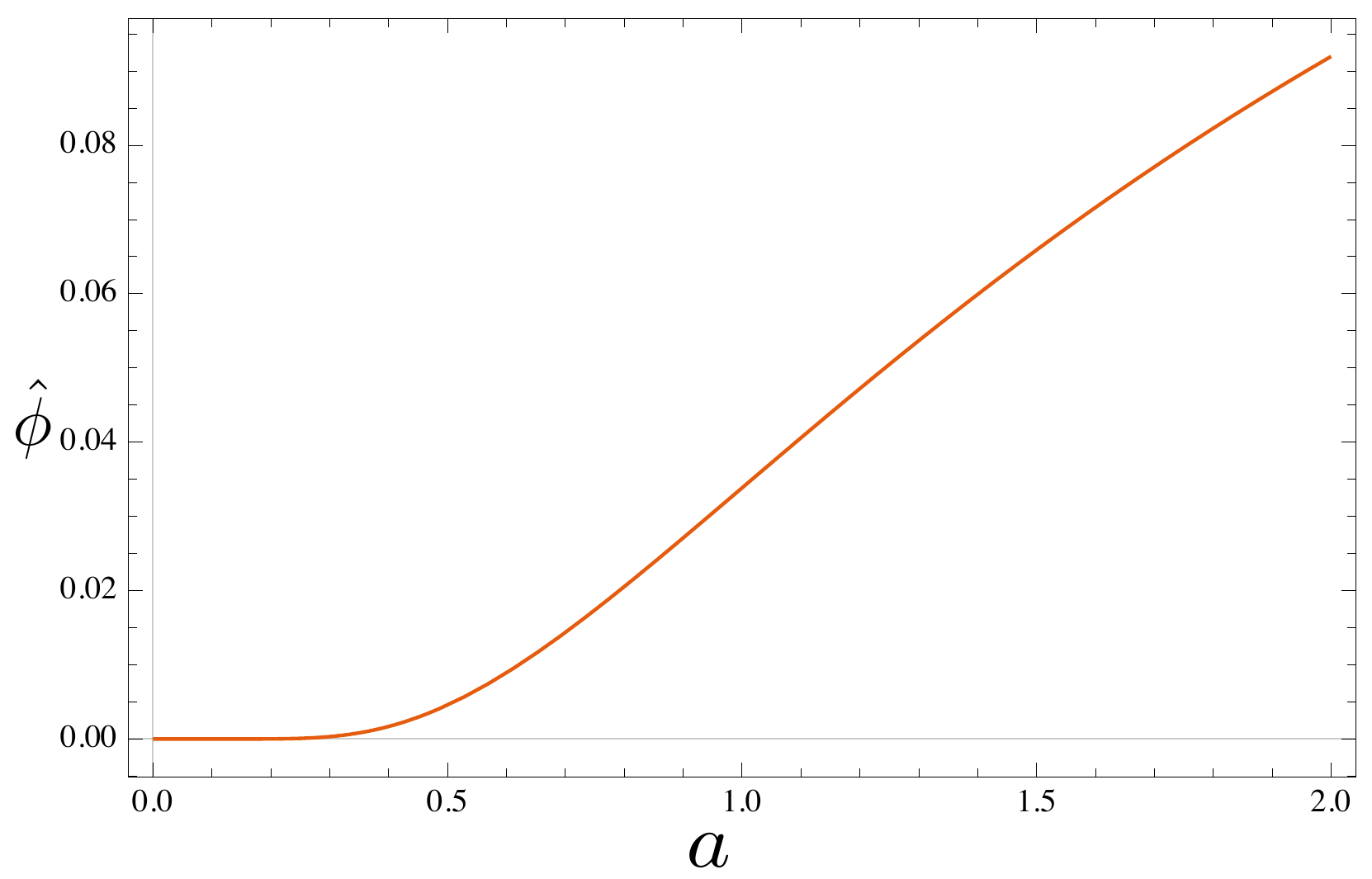}
\includegraphics[scale=0.4]{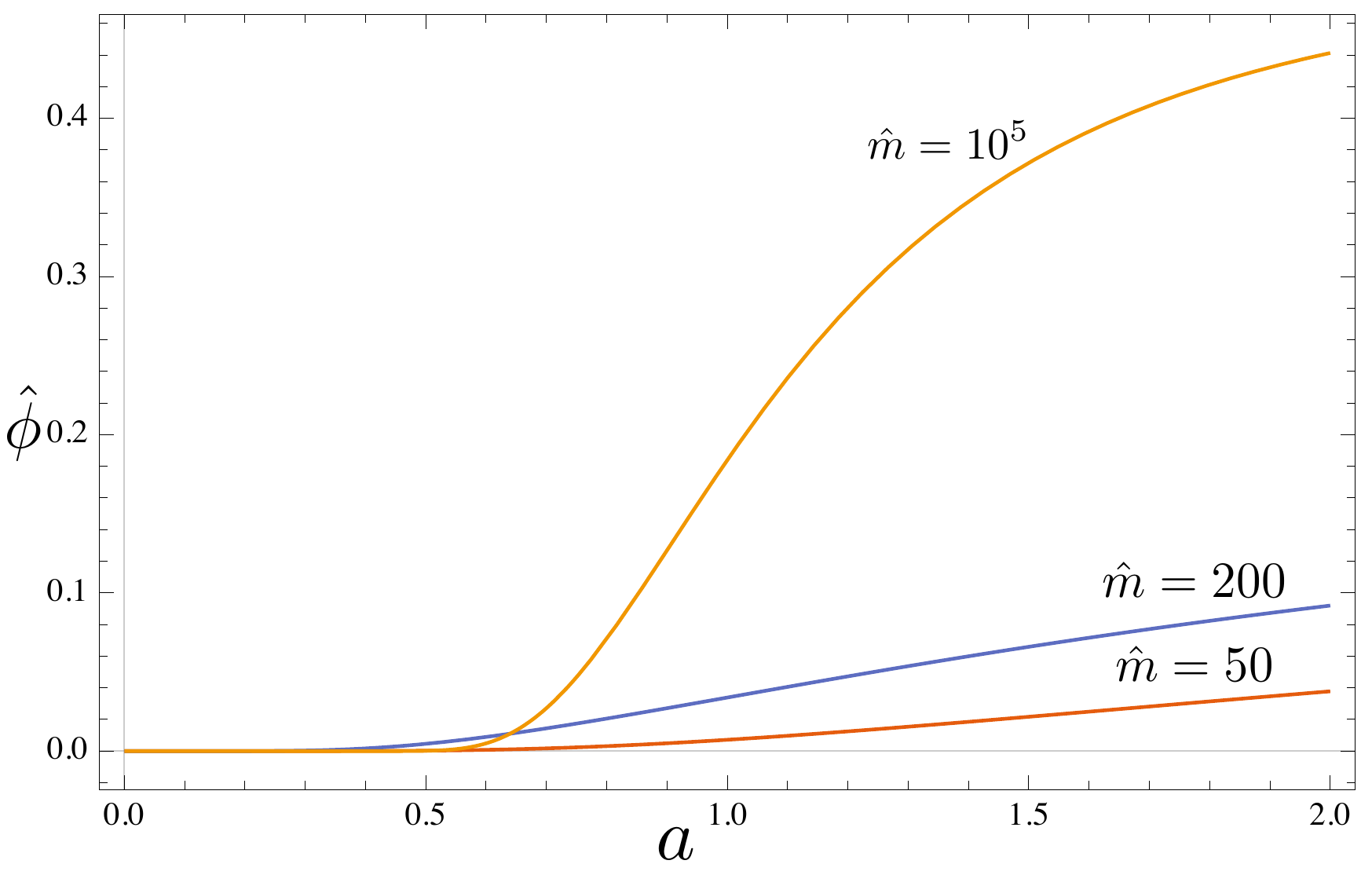}
\caption{In the right panel, numerical solutions for the dilaton field for a
range of values of the dilaton mass for a polytropic neutron star with  $\Gamma = 2$, $\hrho_{\infty} = 10^{-4}$, and $M_0=0.01 M_{\rm Pl} $. In the left panel
the dilaton field with $\Gamma = 2.07$ for a polytropic star  mimics the behaviour of the $\hat{m} = 200$ curve in the right  panel whilst here $M_0=0.01 M_{\rm Pl} $, $\hat m = 100$ and $\hrho_{\infty} = 10^{-4}$.
}
\label{dilstar3}
\end{figure}

We  begin with a constant density star with equation of state $\rho = \rho_0$ and a smoothed out transition to the asymptotic matter distribution surrounding the star. We plot the scalar field profile for a range of central densities $\hrho_c$ in the left panel of figure~\ref{dilstar1} and for a range of effective dilaton masses in the right panel of figure~\ref{dilstar1}. We can see  that for low central densities the field smoothly transitions from its minimum inside the star to the asymptotic minimum far from the stellar surface.
Increasing the central density or the mass of the dilaton have a similar effect on scalar profile. The transition becomes sharper and more kink like
and occurs closer to the surface. The dilaton scalar profiles are however noticeably smoother than the chameleon profiles from the previous section.

For a polytropic neutron star we use the  modified equation of state \ref{eq5.77}. In the panels of figure ~\ref{dilstar3}, we  show the evolution of the scalar field for a range of scalar masses. Again and as for chameleons we show in the left panel that changing $\Gamma$ can be mimicked by a change of the central mass and vice versa, i.e. the equation of state is degenerate with a change of the dilaton parameters.

Finally  in fig.~\ref{chamstar5} we exemplify  the scalar instability that develops in the dilatonic case for $\hrho-3\hP<0$. This criterion therefore appears to be a good model independent way to test the onset of such an instability for stellar structures in both the chameleon and dilaton cases.

\section{Implications and conclusions}

\begin{figure}[!htb]
\centering
\includegraphics[scale=0.4]{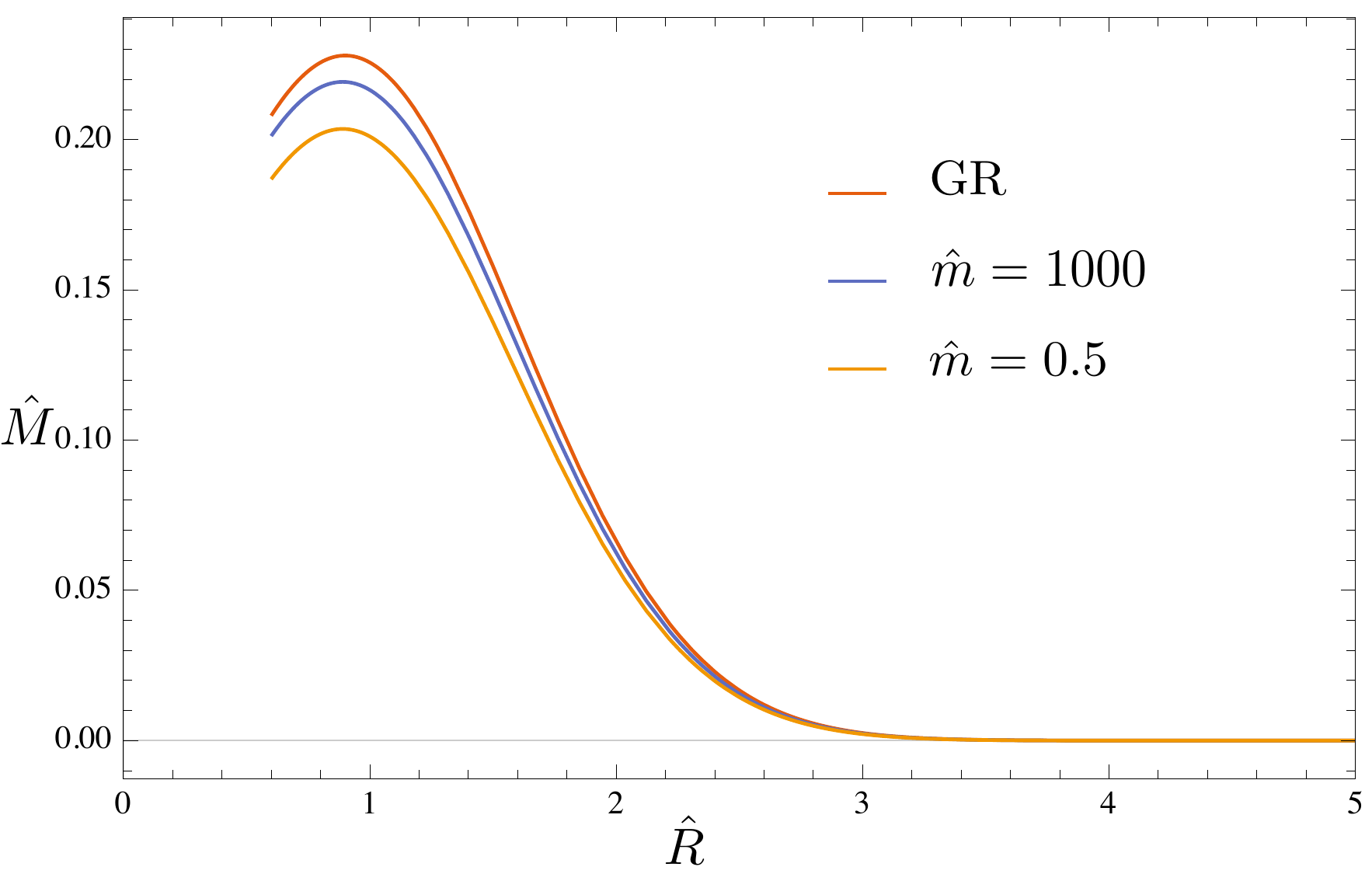}
\includegraphics[scale=0.4]{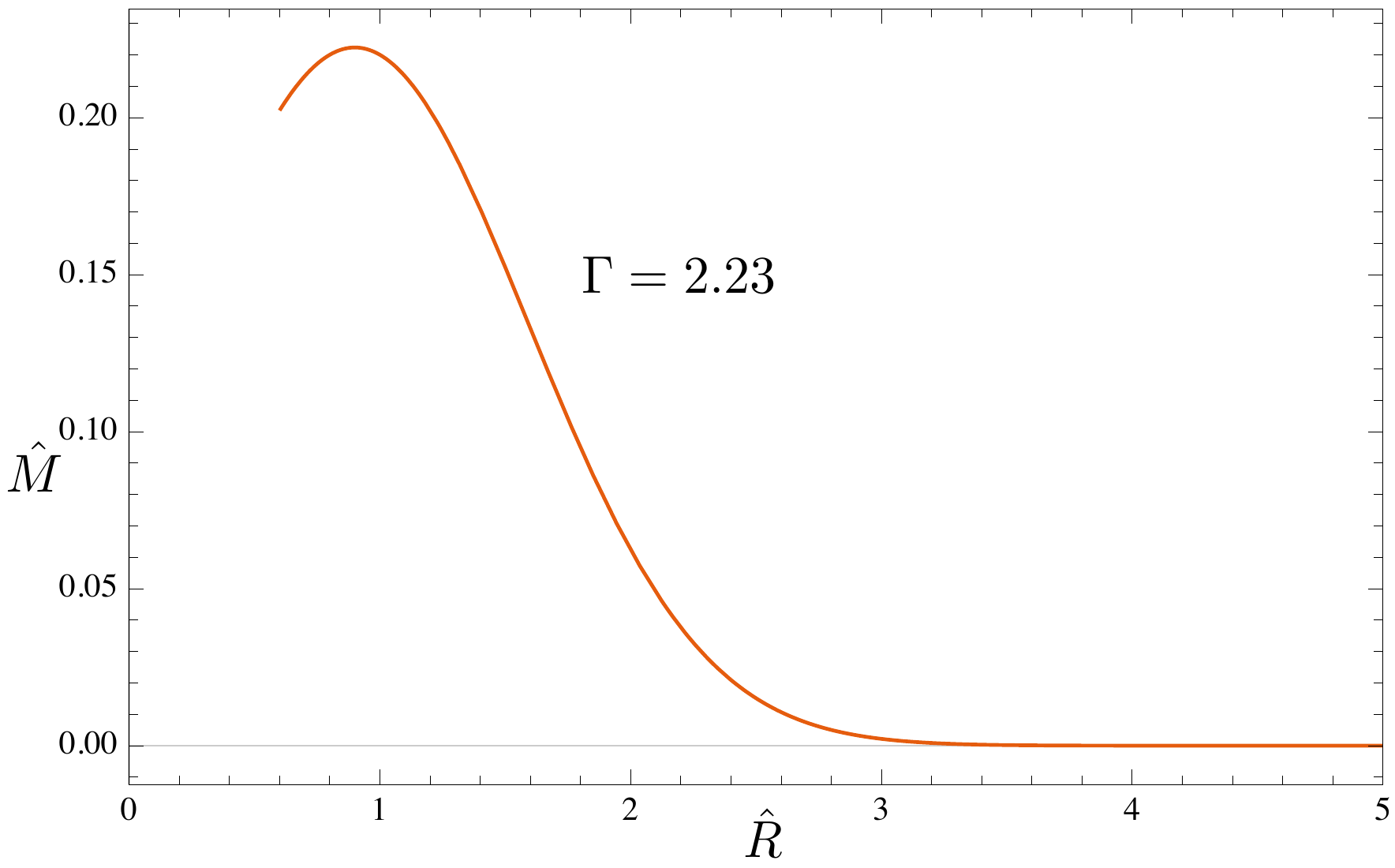}
\caption{In the left panel we show the scaled mass-radius  relationship of a polytropic star ($\Gamma = 2$) for a range of values of the chameleon mass. In the right panel,
the scaled mass-radius  relationship of a polytropic star with $\Gamma = 2.23$ in pure GR. This mimics the $\hat{m} = 1000$ curve of the left panel reasonably well.}
\label{chamstar6}
\end{figure}

We have studied scalar field profiles around neutron stars for  different  modified gravity models corresponding to two screening mechanisms, i.e. the chameleon and Damour-Polyakov mechanisms. This has been done for two classes of models: the inverse power law chameleons and the environmentally dependent dilaton. The latter is particularly interesting as it is expected to arise in the strong coupling regime of string theory. We have solved the dynamics of these models using  a relaxation algorithm allowing us to  study the modified Tolman-Oppenheimer-Volkoff equations for constant density stars and the slightly more realistic polytropic neutron stars for a range of model parameters.

We find non-trivial modifications to the stellar structure due to the addition of a scalar degree of freedom to gravity. In particular we find that deep inside the star, the scalar field settles down to the minimum of the effective potential which depends on the modified gravity model used and the details of the equation of state used to model the neutron star. Close to the stellar surface, the scalar field starts responding to the changing ambient density and this response is either sluggish or extremely rapid depending on the model parameters used, i.e. eventually on the central mass of the scalar field. Finally far away from the neutron star the scalar field settles down to the minimum of the effective potential with the ambient cosmological density. Thus the scalar profile on average looks like a smoothed out kink-like solution extrapolating between the two minima. We find that the scalar field develops an instability deep in the stellar centre for $\trho-3\tP<0$. This criterion appears to be robust for a range of parameter values and for both chameleon and dilaton theories. Indeed a crude theoretical estimate shows that we would expect the radial mode of spherically symmetric scalar perturbations to develop an instability for $\trho-3\tP\lesssim 0$.

\begin{figure}[!htb]
\centering
\includegraphics[scale=0.4]{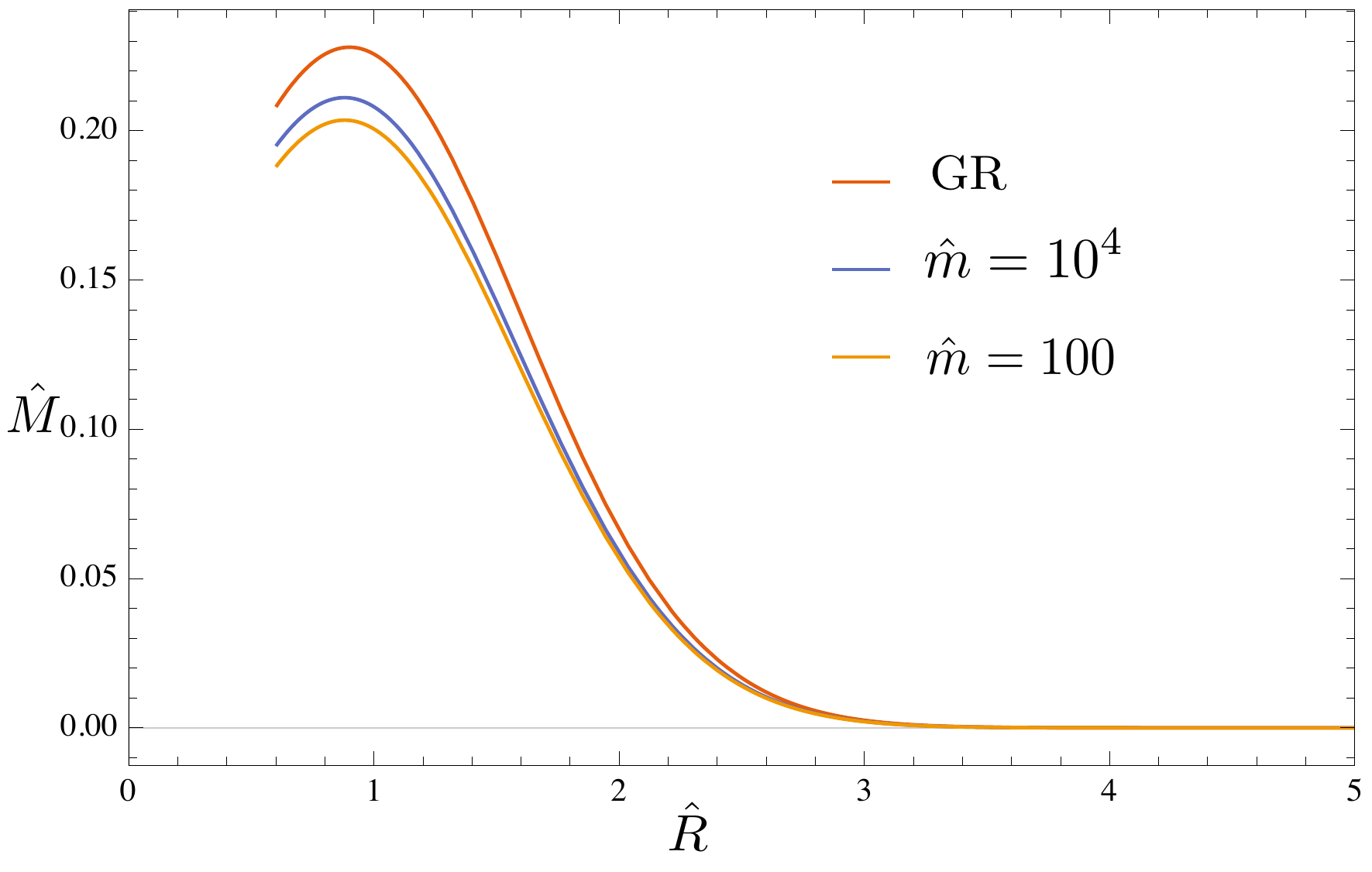}
\includegraphics[scale=0.4]{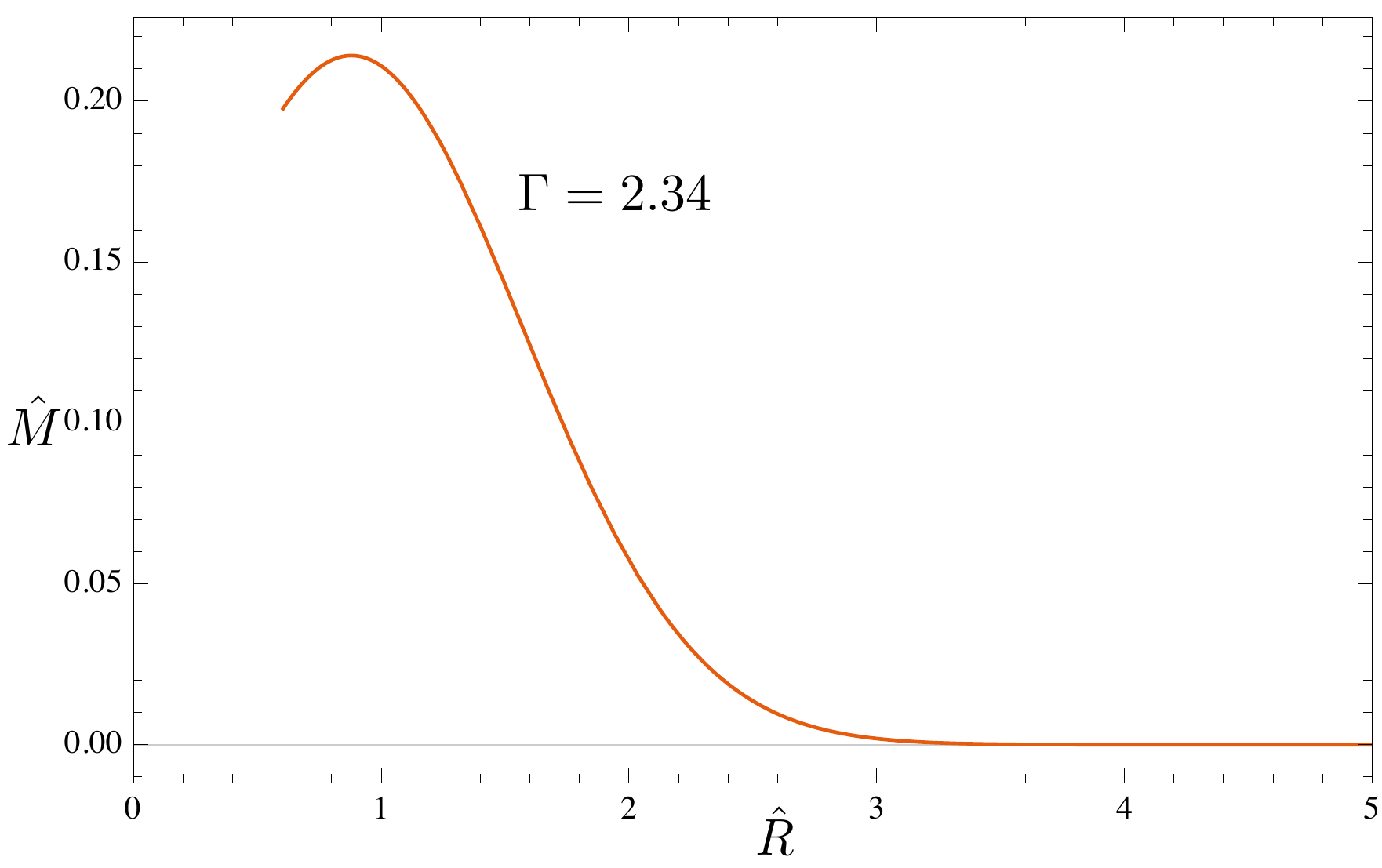}
\caption{In the left panel, we show the scaled mass-radius relationship of a polytropic star ($\Gamma = 2$)  for a range of values of the central dilaton mass. In the right panel, we show the scaled mass-radius relationship of a polytropic star with $\Gamma = 2.34$ in pure GR. This mimics the $\hat{m} = 10^4$ curve in the left panel reasonably well.}
\label{dilstar6}
\end{figure}

 For realistic models which pass the solar system tests, we expect the mass of the scalar field to be very large deep inside the star, due to the very large central density. Hence the kink-like solution should  vary very abruptly when interpolating between the two minima, implying that the effects of the scalar field on the dynamics of star should be reduced. This can be inferred from  the figure \ref{dilstar3} where the profile is depicted and the  figure \ref{dilstar6} where the change of the mass-radius relationship is shown to be smaller when the central mass increases. Of course, taking into account the actual solar system constraints on the models and the resulting effects on the field profile and the mass radius relationship is beyond the scope of the present paper, as it very likely that different numerical methods would have to be used.

Another interesting aspect of the  numerical solutions is the very high degree of degeneracy between parameters of the modified gravity theory and parameters of the neutron star equation of state. We find that by changing the effective mass parameter of chameleons and dilatons we can mimic any change in the scalar profile that results from varying the stellar central density or the polytropic index of the equation of state. This can be strengthened by analysing  the mass-radius relationship for the chameleon, as  in figures \ref{chamstar6},  and for the dilaton, as  in figures \ref{dilstar6}. We have determined the radius $r_\star$ of the star as the point where the pressure vanishes $\tilde P(r_\star)=0$ and the mass as $G_N M_\star= m(r_\star)$. We have used the rescaled mass and radius defined by
\be
\hat M= K^{1/4(\Gamma -1)} M_\star,\ \ \hat R= K^{-1/4(\Gamma -1)} r_\star
\ee
as $K$ has mass dimension $4(1-\Gamma)$, which has the advantage of making the overall normalisation of the polytropic equation of state $K$ not very relevant to our results \cite{cook}.

 It can be seen  that varying the central mass of the scalar for instance has a similar effect on the mass-radius relationship as varying the polytropic index for the stellar equation of state. As the experimental constraints on the stellar equation of state for  neutron stars are weak \cite{Ozel:2009da},  it is likely  that for dilatons and chameleons  the neutron star observables that depend only on the mass and the radius of the star are not guaranteed to distinguish between small differences in  the equation of state versus small modifications to gravity.  It was argued in \cite{Psaltis:2007rv} that observables that depend on the effective surface gravity of the neutron star can in principle break this degeneracy. However it has  been shown that this might not be the case for $f(R)$ theories \cite{Cooney:2009rr}. It may be possible to break such a degeneracy by using the near independence on the equation of state of the relationship between the reduced moment of inertia $\bar I= I/G_N^2 M_\star^3$ and the compactness $C= G_N M_\star /r_\star$ of the star \cite{Breu:2016ufb}. Other observables  might be useful, such as cooling rates of neutron stars. A neutron star cools via photon and neutrino emission.
Whilst the photon luminosity depends only on the density
of the photosphere surrounding the neutron star, neutrino cooling is sensitive to the central temperature. As a result, it would be interesting to analyse if  observations of
the surface temperatures of neutron stars could lead
to useful constraints on the deviations from general relativity \cite{dpage}. Such an analysis for chameleons and dilatons, however, is beyond the scope of this work.
Finally it would be extremely interesting to study how our results change when the equation of state for the matter in the star is modified, i.e. not of the polytropic type anymore. We hope to come back to this point in another publication.

\acknowledgments

We would like to thanks E. Babichev and D. Langlois for suggestions.
RJ is supported by the Cambridge Commonwealth Trust and Trinity College, Cambridge. This project has received funding from the European Union’s Horizon 2020 research and innovation programme under the Marie Skłodowska-Curie grant agreement No 690575. This article is based upon work related to the COST Action CA15117 (CANTATA) supported by COST (European Cooperation in Science and Technology). ACD is supported in part by STFC UK under ST/L000385/1 and ST/L000636/1.

\providecommand{\href}[2]{#2}\begingroup\raggedright\endgroup

\end{document}